\documentclass[journal]{IEEEtran}

\usepackage{epsfig}

\def\um{\ifmmode {\mathrm{\mu m}}\else
                  \textrm{$\mu$m }\fi}%
\def\GeV{\ifmmode {\mathrm{\ Ge\kern -0.1em V}}\else
                   \textrm{Ge\kern -0.1em V}\fi}%
\def\MeV{\ifmmode {\mathrm{\ Me\kern -0.1em V}}\else
                   \textrm{Me\kern -0.1em V}\fi}%
\def\keV{\ifmmode {\mathrm{\ ke\kern -0.1em V}}\else
                   \textrm{ke\kern -0.1em V}\fi}%
\def\eV{\ifmmode  {\mathrm{\ e\kern -0.1em V}}\else
                   \textrm{e\kern -0.1em V}\fi}%
\def\uW{\ifmmode  {\mathrm{\mu  W}}\else
                   \textrm{$\mu$W}\fi}%

\begin{document}


\title {Data production models for the CDF experiment }

\author{  \begin{center}
  J.~Antos,
  M.~Babik,
  D.~Benjamin,
  S.~Cabrera,
  A.W.~Chan,
  Y.C.~Chen,
  M.~Coca,
  B.~Cooper,
  K.~Genser,
  K.~Hatakeyama,
  S.~Hou,
  T.L.~Hsieh,
  B.~Jayatilaka,
  A.C.~Kraan,
  R.~Lysak,
  I.V.~Mandrichenko,
  A.~Robson, 
  M.~Siket,  
  B.~Stelzer,
  J.~Syu,
  P.K.~Teng,
  S.C.~Timm,
  T.~Tomura,
  E.~Vataga,
  S.A.~Wolbers, and 
  P.~Yeh  \end{center}
\thanks{ J.~Antos, and M.~Babik are with 
         Institute of Experimental Physics, Slovak Academy of Sciences, Slovak Republic.}
\thanks{ D.~Benjamin, S.~Cabrera, and M.~Coca are with Duke University, Durham, NC 27708, USA. }
\thanks{ A.W.~Chan, Y.C.~Chen, S.~Hou, T.L.~Hsieh, R.~Lysak, M.~Siket, P.K.~Teng,
         and P.~Yeh are with 
         Institute of Physics, Academia Sinica, Nankang, Taipei, Taiwan.}
\thanks{ B.~Cooper is with University College London, London WC1E 6BT, United Kingdom. }
\thanks{ K.~Genser, I.V.~Mandrichenko, J.~Syu, S.C.~Timm, and S.A.~Wolbers are with
         Fermi National Accelerator Laboratory, Batavia, IL 60510 USA.}
\thanks{ K.~Hatakeyama is with The Rockefeller University, New York, NY 10021, USA. }
\thanks{ B.~Jayatilaka is with University of Michigan, Ann Arbor, MI 48109, USA. }
\thanks{ A.C.~Kraan is with University of Pennsylvania, Philadelphia, PA 19104, USA. }
\thanks{ A.~Robson is with Glasgow University, Glasgow G12 8QQ, United Kingdom. }
\thanks{ B.~Stelzer is with University of California, Los Angeles, Los Angeles, CA 90024, USA. }
\thanks{ T.~Tomura is with University of Tsukuba, Tsukuba, Ibaraki 305, Japan. }
\thanks{ E.~Vataga is with University of New Mexico, Albuquerque, NM 87131, USA. }
}
\maketitle

\begin{abstract}

The data production for the CDF experiment is conducted on
a large Linux PC farm designed to meet the needs of data collection 
at a maximum rate of 40 MByte/sec.
We present two data production models that exploits 
advances in computing and communication technology.
The first production farm is a centralized system that
has achieved a stable data processing rate of approximately 2 TByte 
per day.
The recently upgraded farm is migrated to the SAM (Sequential Access 
to data via Metadata) data handling system.
The software and hardware of the CDF production farms has
been successful in providing large computing and data throughput 
capacity to the experiment.

\end{abstract}

\section{Introduction}

The Collider Detector at Fermilab (CDF) detector is a large 
general purpose detector for studying proton-anti-proton 
collisions at the Fermilab Tevatron Collider.
The CDF detector has been upgraded to take advantage of the 
improvements in the accelerator~\cite{CDF2}.  
Computing systems were also upgraded for processing larger 
volumes of data collected in Run II since 2000.
The type of data processing required for CDF is a decoupled parallel 
processing of ``events", where each event is a
detector measurement of a beam collision.
A hardware and software trigger system is used to
store and save data from interesting collisions.
The events are saved in ``raw'' data format.
On the farm each event is processed through a CPU intensive 
reconstruction program that transforms digitized electronic signals 
from the CDF sub-detectors 
into information that can be used for physics analysis. 
The quantities calculated include particle trajectories and momentum, 
vertex position, energy deposition, and particle identities.

The production farms are collections of dual CPU PCs running Linux,
interconnected with 100 Mbit and gigabit ethernet.
The challenge in building and operating PC farms is in 
managing the large flow of data through the computing units.
The control software is required to be precise on bookkeeping
for having every raw data file processed and the output
stored in run sequence.
Hardware and program errors do occur, therefore easy intervention
and recovery are also required.

In this paper we describe the hardware integration and software for
operation of the CDF production farms. 
We first describe the requirements and design goals 
of the system.  
The first farm's hardware and software control is described.
The performance and experiences with this system is presented.
The upgrade aims for data processing in a distributed computing 
environment.
Software control is migrated using the Fermilab developed "Sequential 
Access via Metadata" (SAM) system \cite{SAM} for data handling.
Performance of data production with the SAM production farm are 
also presented.

\section{Requirements}

To achieve the physics goals of the CDF experiment at the Fermilab
Tevatron, the production computing system is required to process the
data collected by the experiment in a timely fashion.
In 2001 through 2004 the CDF experiment collected a maximum 
of 75 events/second at a peak throughput of 20 MByte/sec.
The recent upgrade has improved the bandwidth to 40 MByte/sec.
Raw data are collected in parallel in eight data streams.
Events of a similar type are collected into 1 GByte files 
for a data collection period assigned with a unique run number.
The output of event reconstruction is split into many physics data-sets,
placing similar physics data together on disk or tape files for 
faster and more efficient physics analysis.
The output event size is approximately the same as the input.
Therefore the system output capacity is also required to be 
approximately 40 MByte/sec.

To accomplish rapid data processing through the farms, adequate 
capacity in network and CPU is required.
The event processing requires 2-5 CPU seconds on a Pentium III 1 GHz PC.
The exact number depends on the type of event, the version of the 
reconstruction code, and the environment of the collision. 
These numbers lead to requirements of the equivalent of about 500
Pentium III 1 GHz CPUs, assuming 100\% utilization of the CPUs. 

The production farm operation is required to be
easily manageable, fault-tolerant, scalable, with good monitoring and 
diagnostics.
Hardware and software options were explored to meet the
requirements for the system.  These include large 
symmetric multiprocessing systems,
commercial UNIX workstations, and alternative network configurations.
Prototype systems were built and tested before the final design was
chosen and production systems built.

\section{Architecture }

\begin{figure}[t!]
  \centering
  \epsfig{file=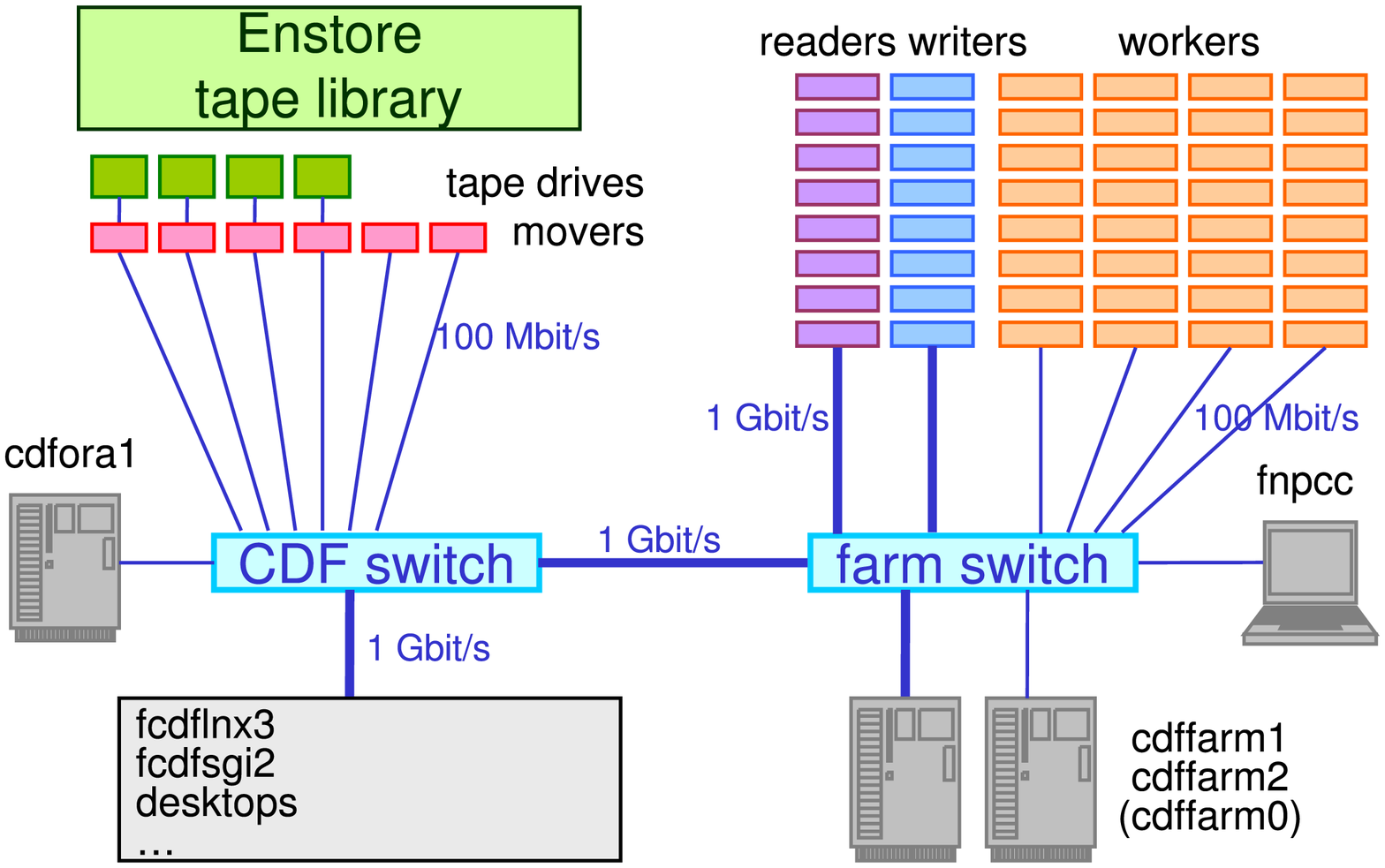,width=0.95\linewidth}
  \caption{CDF production farm architecture.
  \label{fig:farm} }
  \vspace{-.5cm}
\end{figure}

The CDF data production farm consists of a large number
of PCs (workers) that run the CPU-intensive codes,
PCs (readers and writers) that buffer data
into and out of the farm and servers providing various services.
The hardware architecture is shown in Fig.~\ref{fig:farm}.
It has two server nodes {\sf cdffarm1} and {\sf cdffarm2}.
{\sf cdffarm1} is a SGI O2000 machine 
that host a batch submission system and a database server.
{\sf cdffarm2} is a dual Pentium server running control daemons 
for resource management and job submission.
Monitoring and control interfaces for farm operation includes a 
java server to the control daemons and a web server for monitoring.
The disk space uses a Fermilab developed ``dfarm'' file system \cite{dfarm}.
It is a distributed logical file system using a collection of
IDE hard-disks of all dual Pentium nodes.
The job scheduling on the production farm is controlled by a batch 
management system called FBSNG developed by the Computing Division 
at Fermilab \cite{FBS}.
The CDF Data Handling system is a well-defined interface 
\cite{data_handl} to a mass storage system 
of a pByte Enstore tape library \cite{Enstore}. 

The dual Pentium nodes were purchased over many years.
Old nodes were replaced after three years in service.
At its peak in mid-2004, there were 192 nodes in service.
The dfarm capacity of the collected worker hard-disks was as
large as 23 TByte including three
file servers each having 2 TByte.
The IDE hard-disk size varies from 40 to 250 GByte.

The input and output (I/O) nodes are configured to full
capacity in data through-put to the farm.
A total of 16 nodes equipped with optical giga-links
are configured with the {\sf pnfs} file system \cite{pnfs}
for access to the Enstore storage.
A 48 port Cisco switch module was added recently to provide gigabit
ethernet over copper switching.
Additional I/O nodes may be added if needed.
The number of workers can be scaled to as large a number as is required.
However, the total data through-put capacity to
Enstore storage is limited by the number of Enstore movers 
(tape-drives) available.

\begin{table}[t!]
  \begin{center}
  \begin{tabular}{|c|c|c|c|c|} \hline
   Stream  
   & data-sets & events/GByte & total event (\%) & total size (\%) \\
   \hline
   A & aphysr &  2720      &  3.8  & 7.7 \\
   B & bphysr &  5470      &  9.9  & 5.5 \\
   C & cphysr &  6770      &  9.2  & 7.5 \\
   D & dphysr &  2570      &  3.7  & 7.9 \\
   E & ephysr &  5930      &  17.0 & 15.7 \\
   G & gphysr &  6140      &  26.4 & 23.5 \\
   H & hphysr &  6050      &  19.6 & 17.7 \\
   J & jphysr &  5520      &  10.3 & 10.3 \\ \hline
  \end{tabular}
  \caption{Statistics of data streams of a typical run taken in 
     June 2004 containing all sub-detectors. 
     The raw data files are 1 GByte in size.
     Listed are the number of events per GByte,
     ratio of total events and total file size.
  \label{tab:rawdata} }
  \end{center}
  \vspace{-.5cm}
\end{table}

\begin{figure}[b!]
  \centering\epsfig{file=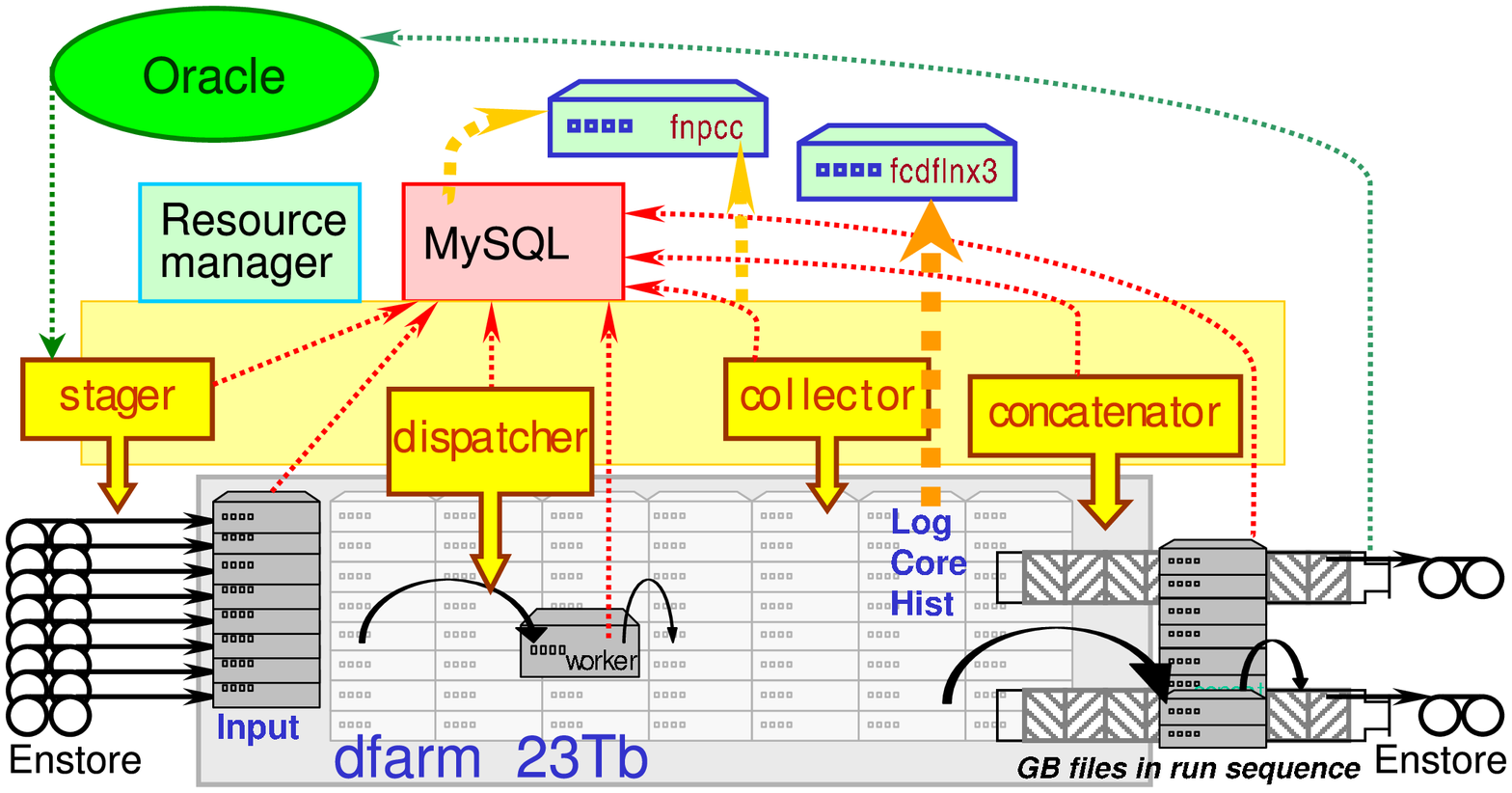,width=1.\linewidth}
  \caption{Flow control in the FPS production farm.
  \label{fig:flow} }
  \vspace{.5cm}
\end{figure}

\section{Farm Processing System }

Raw data from the experiment is first written to tape 
in the Enstore mass storage system.
Statistics of the eight data streams of a typical run
are listed in Table~\ref{tab:rawdata}.
These tapes are cataloged in the CDF Data File Catalog (DFC) \cite{DFC}
as a set of tables in an Oracle database 
(accessed via {\sf cdfora1} in Fig.~\ref{fig:farm}).
After the data is written to 
tape and properly cataloged, and once the necessary calibration 
constants exist, the data is available for reconstruction on the farms.

The production farm is logically a long pipeline with the constraint 
that files must be written to mass storage in order.
The input is fetched directly from Enstore tapes and 
the output is written to output tapes.
The data flow is illustrated in Fig.~\ref{fig:flow}
for the files moving through dfarm storage controlled by four
production daemons.  The daemons communicate with the
resource manager daemon and the internal database to schedule
job submission.
The internal database is a MySQL \cite{MySQL} system used for 
task control, file-tracking, and process and file history.
The DFC records are fetched at the beginning of staging input data.
Output files written to tapes are recorded in the DFC. 
Job log files and other logs and files are collected to the user 
accessible node ({\sf fcdflnx3}). 
Operation status is monitored by a web server ({\sf fnpcc}).

The operation daemons are configured specifically for 
production of a input ``data-set''.
For raw data, each data stream is a data-set.
The input files are sent to worker nodes for reconstruction.
Each worker node (dual-CPU) is configured to run two 
reconstruction jobs independently. An input file is 
approximately 1 GByte in size and is expected to run for about 
5 hours on a Pentium III 1 GHz machine.
The output is split into multiple files, with each file corresponding
to a data-set defined by the event type in the trigger system.
An event may satisfy several trigger patterns and be
written to multiple data-sets that are consistent with that event's 
triggers. 
Each data-set is a self-contained sample for physics analysis. 
The total number of output data-sets is 43 
in the most recent trigger table.

\begin{figure}[t!]
  \centering\epsfig{file=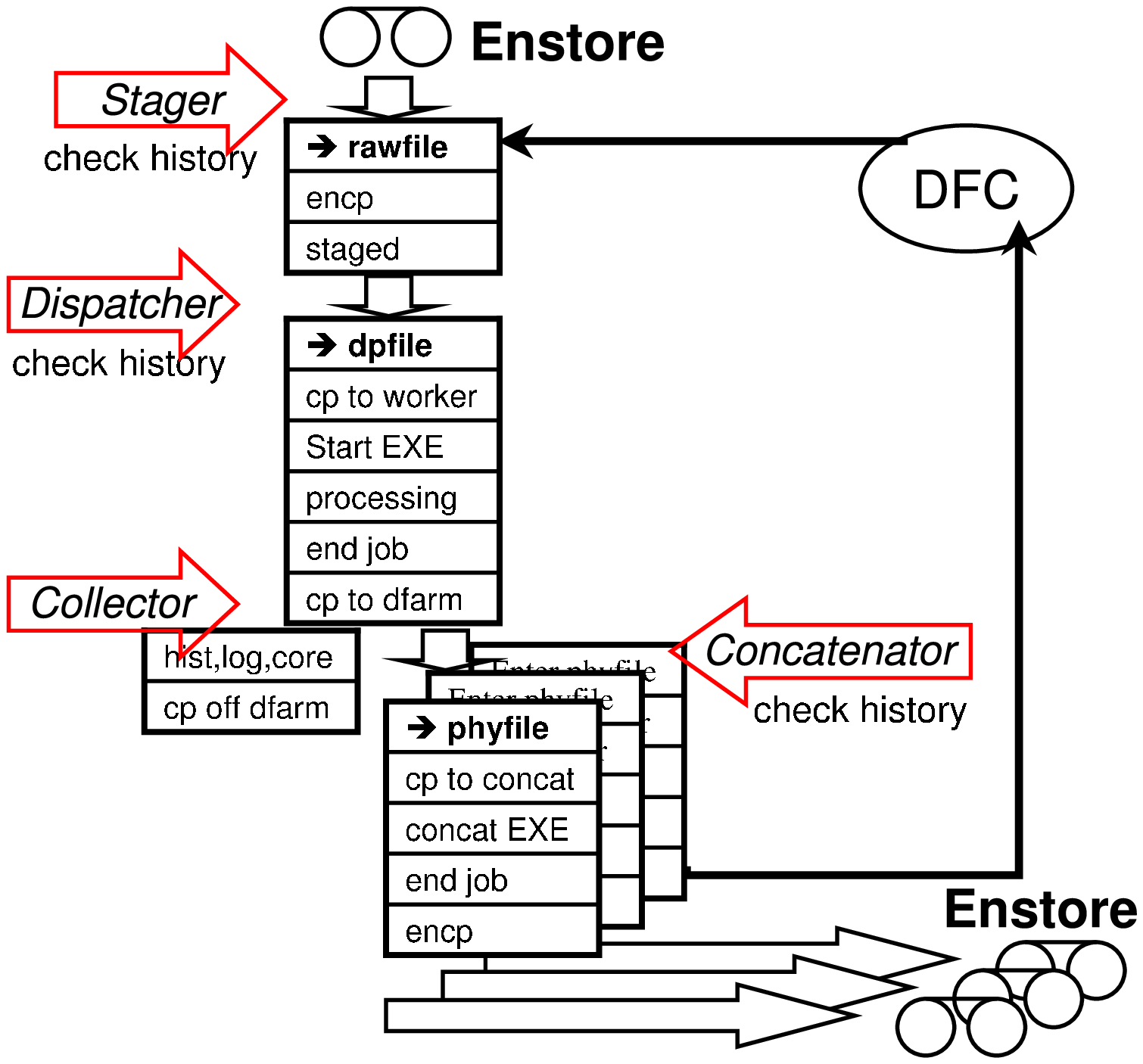,width=.75\linewidth}
  \vspace{-.5cm}
  \caption{Task control for a farmlet. Status is recorded for each
  input file in MySQL database.
  \label{fig:bookkeeping} }
\end{figure}


The Farm Processing System (FPS) is the software that
manages, controls and monitors the production farm.  
It is flexible and allows configuration
for production of data-sets operated independently 
in parallel farmlets.
A farmlet contains a subset of the farm resources
specified for the input data-set, the executable 
and the output configuration for concatenation merging
small output files into large ones.
Its execution is handled by its own daemons taking care of 
consecutive processing in production and its records are written 
in the internal database.  The task control by FPS for
a farmlet is illustrated in Fig.~\ref{fig:flow} and
Fig.~\ref{fig:bookkeeping}.
The daemons of the farmlets are :

\begin{itemize}
   \item {\bf Stager} is a daemon that is responsible for finding and 
   delivering data from tapes based on user selection for a set of 
   data files or run range in the data-set. 
   Jobs are typically submitted one ``file-set'' at a time.
   A file-set is a collection of files with a typical size of 10 GByte.
   The stager fetches DFC records for input and checks that
   proper calibration constants are available.
   The staging jobs are submitted to the input I/O nodes 
   and the file-sets are copied to their scratch area, 
   and afterward to dfarm.

   \item {\bf Dispatcher} submits jobs through the batch manager to 
   the worker nodes and controls their execution. 
   It looks for the staged input file, which is then
   copied into the worker scratch area.
   The binary ``tarball'' (an archive of files created with the Unix
   tar utility) containing the executable, complete libraries,
   and control parameter files (in TCL language) are also copied. 
   This allows the reconstruction program to run locally on the 
   worker nodes and the output files, of various sizes from 5 MByte 
   to 1 GByte, are written locally.
   At the end of the job the output files are then copied back to dfarm.
   In case of abnormal system failure, job recovery is performed
   and the job is resubmitted.

   The ``tarball'' is self contained and is suitable for distributed computing
   running on compatible Linux clusters.

   \item {\bf Collector} gathers any histogram files, log files and any 
   additional relevant files to a place where members of the 
   collaboration can easily access them for validation or 
   monitoring purposes.

   \item {\bf Concatenator} writes the output data that is produced 
   to the selected device (typically the Enstore tape) in a timely 
   organized fashion. 
   It checks the internal database records for a list of files to be 
   concatenated into larger files with a target file size of 1 GByte.
   It performs a similar task as the dispatcher, with 
   concatenation jobs submitted to output nodes.
   The output nodes collect files corresponding to a file-set size
   ($\approx 10$ GByte) from dfarm to the local scratch area and 
   executes a merging program to read events in the input files
   in increasing order of run numbers.
   It has a single output truncated into 1 GByte files.
   These files are directly copied to tapes and DFC records are written.
\end{itemize}

Since all of the farmlets share the same sets of computers and
data storage of the farm, the resource management is
a vital function of FPS for distribution and prioritization 
of CPU and dfarm space among the farmlets. 
The additional daemons are:
\begin{itemize}
   \item {\bf Resource manager} controls and grants allocations for 
   network transfers, disk allocations, CPU and tape access based on a 
   sharing algorithm that grants resources to each individual farmlet 
   and shares resources based on priorities.  This management of 
   resources is needed in order to prevent congestion either on the 
   network or on the computers themselves and to use resources
   more effectively.
        
   \item {\bf Dfarm inventory manager} controls usage of the 
   distributed disk cache on the worker nodes that serves as a 
   front-end and output cache between the tape pool and the Farm.  

   \item {\bf Fstatus} is a daemon that checks periodically whether 
   all of the services that are needed for the proper functioning of 
   the CDF production farm are available and to check the status 
   of each computer in the farm. 
   Errors are recognized by this daemon and are reported 
   either to the internal database which can be viewed on the web or 
   through the user interfaces in real time.
\end{itemize}

The system control framework of FPS is primarily coded in 
python language \cite{python}.
It runs on one of the server computers ({\sf cdffarm2}) and depends on 
the kernel services
provided by {\sf cdffarm1}, namely the FBSNG batch system, 
the FIPC (Farm Interprocess communication) between the daemons and 
the dfarm server governing available disk space on the worker nodes.  
Daemons have many interfacing components that allow them to communicate
with the other needed parts of the offline architecture of the CDF 
experiment. 
Those include mainly the DFC and the Calibration Database. 


The FPS status is shown in real time on
a web page that gives
the status of data processing, flow of data, and other useful 
information about the farm and data processing.
The web page is hosted on a dual Pentium node 
({\sf fnpcc} on Fig.~\ref{fig:farm}) connected to the farm switch.
The web interface was coded in the PHP language \cite{php}
and RRDtool \cite{rrd} for efficient storage and display 
of time series plots.
The structural elements in the schema include
output from each FPS modules, a parser layer that transforms data into 
a format suitable for RRDtool, a RRDtool cache that stores this data in 
a compact way, and finally the web access to RRD files and queries 
from MySQL for real time display of production information.

The java control interface was designed for platform independent 
access to production farm control using an internet browser.
Information transfer
between the client and server over the network is done using IIOP 
(Internet Inter-ORB protocol) which is part of CORBA \cite{corba}.
It has proved to be stable, and there have been no problems with 
short term disconnections and re-connections. 
An XML processor \cite{xml} is used to generate and interpret the 
internal representation of data. Abstract internal representation 
of data is important to cope with changes in the FPS system.
A Java programming language, Java Web Start technology \cite{webs} 
was used for implementation of a platform independent client.

\section{Bookkeeping }

The control software is required to schedule every
input file with output files to be stored once only
in run sequence.
With hundreds of files being processed at the same time it is 
important to track the status of each file in the farm.
File-tracking and bookkeeping by FPS are recorded
on a MySQL database. 
The database stores information about each individual file, process 
and the history of earlier processing.
Three tables are implemented for each farmlet: stage-in of input
files; reconstruction and output files; and concatenation.
The processing steps tracked by the book-keeping and records in each 
table are illustrated in Fig.~\ref{fig:bookkeeping}. 
Once a file is successfully processed,
its records are copied over to the corresponding history tables.
The file status is used to control the flow of data and 
to make sure that files are not skipped or processed more than once. 
The MySQL database also includes detailed information about the 
status of each file at every point as it passes through the system.
This information is available through the web interface.
This database server was designed to serve thousands of 
simultaneous connections. 

With the help of information
that is stored in the internal database, the system is able in most 
cases to recover and return to the previously known state from 
which it can safely continue to operate.
The daemons checking the file history in the database are not 
instrumented to detect an abnormal failure for a job 
or a file lost due to network or hardware problems.
The concatenator often has to wait for an output file in order to 
combine files in order.
This bottleneck can be a serious problem and is a major consideration 
for relaxing strict ordering file to improve overall 
system performance.

\section{Data processing capacity}  

The FPS farm capacity is described for a major reprocessing 
of all CDF data in March 2004.
The production farm was operated
at full capacity for a six week period.
The CPU speed and data through-put rate are the factors that determine 
the data reconstruction capacity of the production farm.
The computing time required for an event depends on the 
event characteristics determined by the event trigger
in different data streams. 
In addition, the intensity of the proton and antiproton beams matters.
More intense beams lead to multiple events per beam
crossing which in turn lead to more CPU time per event.  
The event size increases with beam intensity from 140 to 180 kByte.
The CPU time per event in reconstruction 
on a dual Pentium III 1 GHz machine (CDF software version 5.3.1)
is around 2 seconds and increases with beam intensity and
event size.

\begin{figure}[b!]
  \centering\epsfig{file=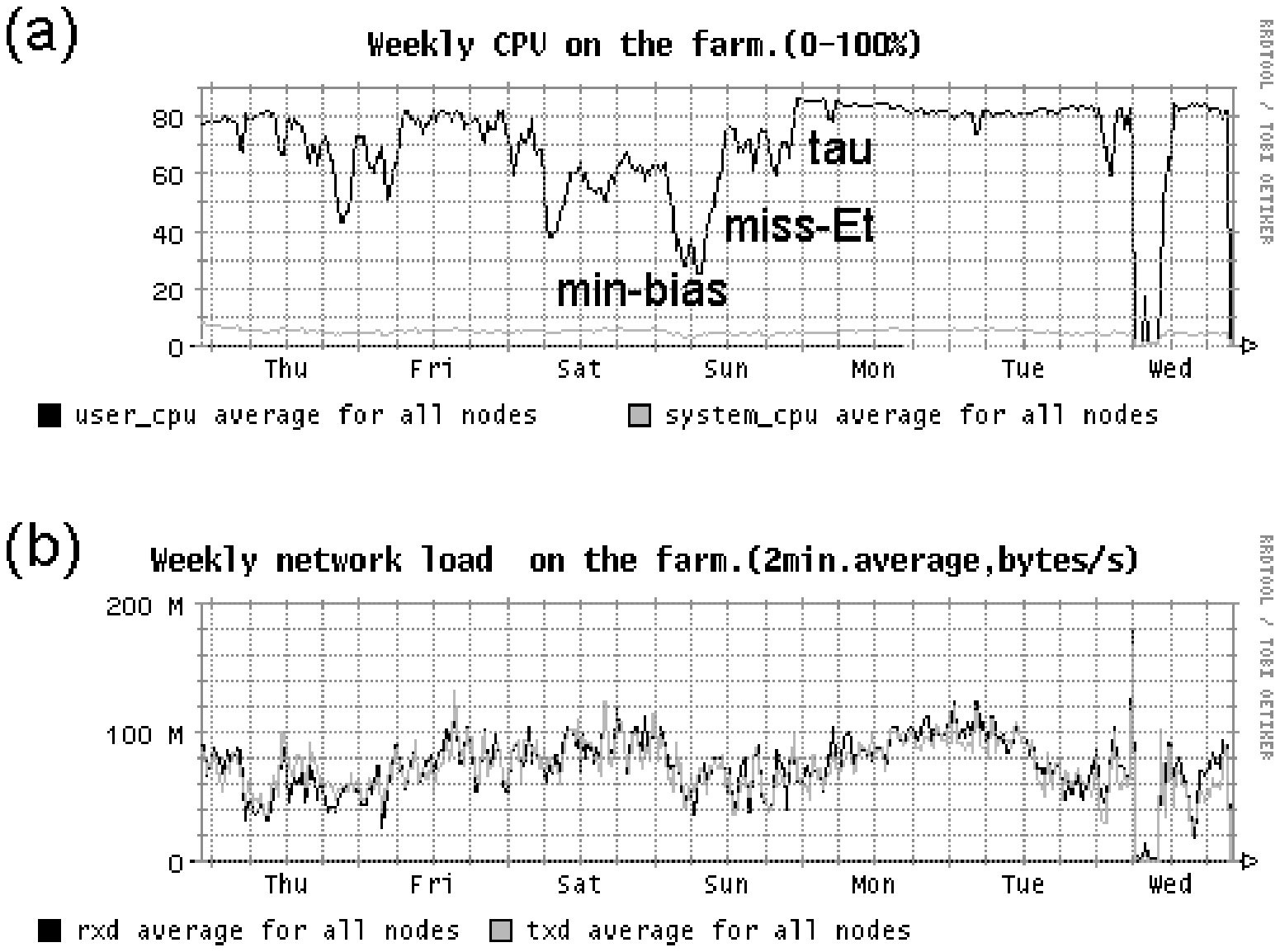,width=1.0\linewidth}
  \caption{(a) CPU load and (b) dfarm traffic 
      of the week of March 18-25, 2004.
  \label{fig:week_cpu} }
\end{figure}

Inefficiency in utilizing CPU comes from the file transfer of
the executable and data files to and from the worker scratch area.
The input data files are staged from Enstore tapes.
The rate of staging data depends on how fast the link to Enstore 
movers is established.
Once a mover is allocated, staging a file-set of 10 GByte takes about 
20 minutes. The data transmission rate varies file by file,
the commonly observed rate is around 10 MByte/sec.

The output of concatenated files are copied to tapes.
The effectiveness in staging data to a tape is a concern because of
the limited dfarm space and output bandwidth.
A concatenation job on the output node collects files of a data-set
with close to 10 GByte at a speed that may reach the maximum IDE 
disk transfer speed of 40 MByte/sec.
It takes an average 10 minutes to copy all the files requested.
The concatenation program reads the numerous small files and writes
output that is split into 1 GByte files. 
On a Pentium 2.6 GHz node the CPU time is about 24 minutes for 
processing 10 GByte.
The job continues by copying the output to Enstore at an average
rate of close to 20 MByte/sec.  
It takes about 10 minutes for writing 10 GByte.
Further delays may be caused by having more than one job accessing the
same hard disk in dfarm, or waiting to write to the same physical tape.

The tape writing is limited to one mover per data-set at a time, 
to ensure that files are written sequentially on tape.
A tape is restricted to files of the same data-set.
The instantaneous tape writing rate is 30 MByte/sec.
However, the average rate drops to below 20 MByte/sec because of 
latency in establishing connection to the mass storage system (this
includes mounting and positioning the tape and establishing the end-to-end
communication).
Running only one data-set on the farm limits the capability of the farm.
Running a mix of jobs from different data-sets in parallel increases 
the through-put of the farm by increasing the output data rate.

\begin{figure}[t!]
  \hspace{.5cm}\epsfig{file=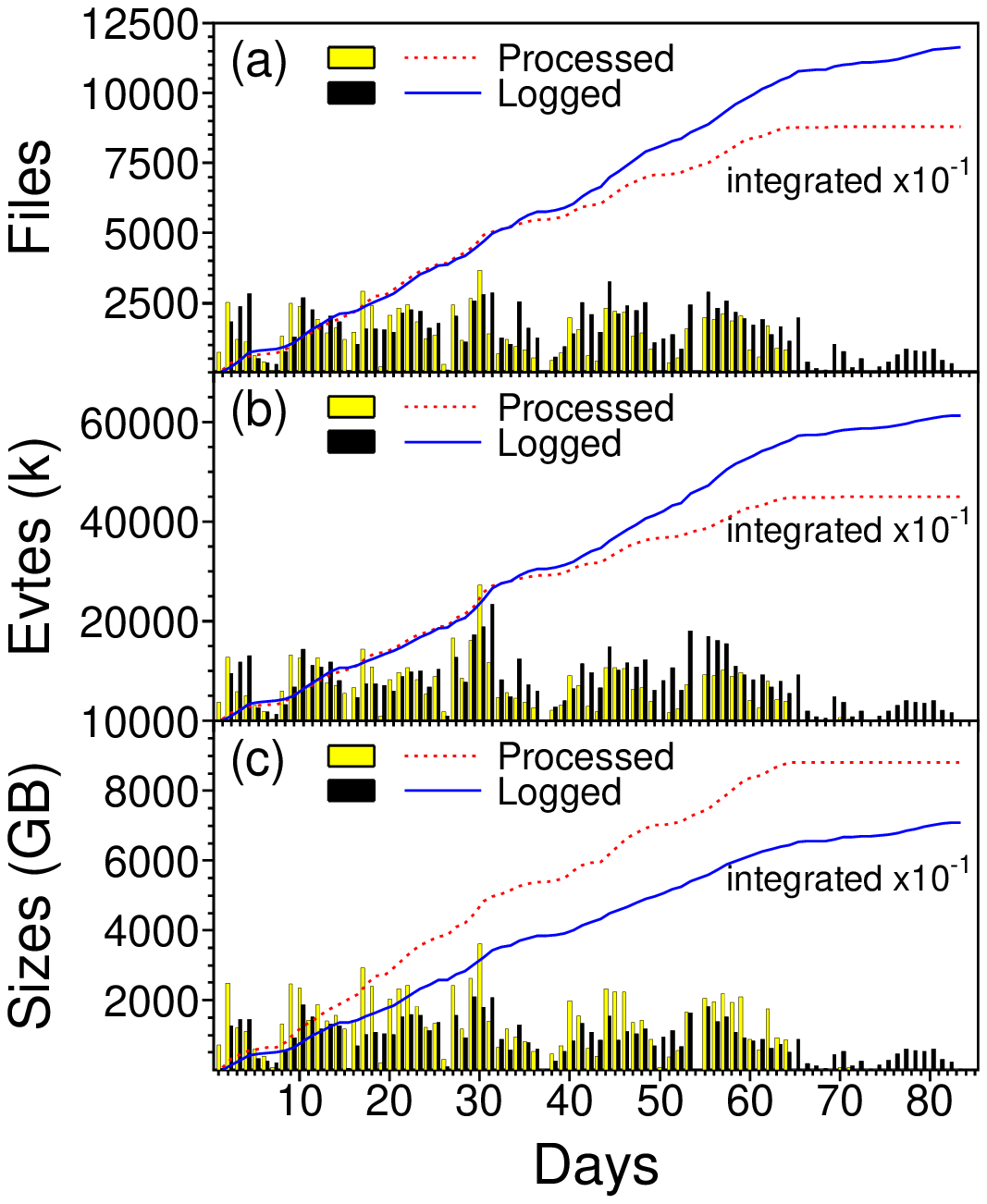,width=1.0\linewidth}
  \vspace{-.5cm}
  \caption{Daily processing rates are shown in histograms for 
    (a) number of files, (b) number of events, and
    (c) data size.  The integrated rates are shown in lines.
    Compressed outputs were created for selected data-sets 
    (about a quarter of the total).
    Event size is reduced by about 30\% and thus a net reduction in 
    output storage.
  \label{fig:stat531} }
\end{figure}

To maximize the farm efficiency the data reprocessing was performed 
on five farmlets with each farmlet processing one data-set.
The tapes were loaded one data-set at a time, therefore
farm CPU usage came in waves shared by a couple data-sets at a time.
The CPU usage for the week of March 18 is shown 
in Fig.~\ref{fig:week_cpu}.
A lag in CPU utilization was observed when the farm switched 
to a new data-set,
seen as the dips in CPU in Fig.~\ref{fig:week_cpu}.a,
because of lack of input files.
File-sets are distributed almost in sequence on a tape
The lag at the beginning of staging in a data-set is because the files 
requested are stored on the same tape, causing
all the stage-in jobs to wait for one tape.
Overall the stage-in is effective in feeding data files to dfarm.
The CPU usage varies for data-sets.
The ``minimum bias'' data-set has smaller event sizes and
the CPU per event is about 40\% less than the average.
When this data-set was processed, the stage-in rate was not 
able to keep up with the CPU consumption.

The output data logging rate is shown in Fig.~\ref{fig:stat531} for
the number of files, number of events, and total file size written 
to Enstore tapes.
Compressed outputs were also created for selected data-sets.
Therefore the total events in output is larger than input by about 25\%.
The event size is reduced and resulted to a net reduction in 
storage by about 20\%.
On average we had a through-put of over 2 TByte (10 million events) 
per day to the Enstore storage.
The data logging lasted two extra weeks for a large B physics data-set
that accounted for about 20\% of the total CDF data. 
It was the latest data-set processed and the tape logging rate 
was saturated at about 800 GByte per day.

\begin{figure}[b!]
  \centering\epsfig{file=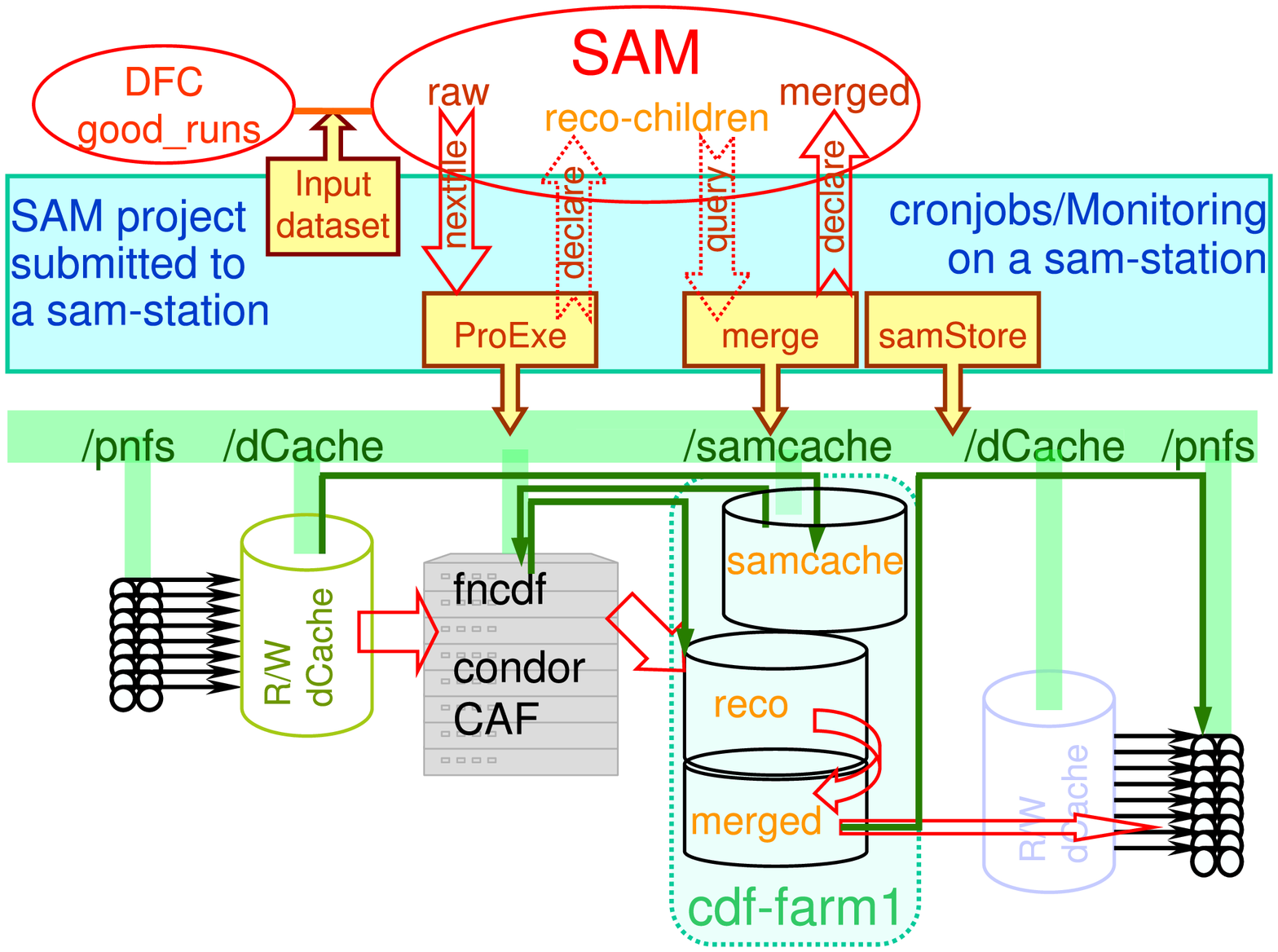,width=1.\linewidth}
  \caption{Data flow and control of production on a SAM farm. 
     Data are transported by SAM to a file cache accessible
     to the Condor CAF facility. Output is sent to a durable storage
     where concatenation is operated.  Merged outputs are 
     declared to SAM and stored to Enstore. 
  \label{fig:samfarm} }
\end{figure}

\begin{figure}[t!]
  \centering\epsfig{file=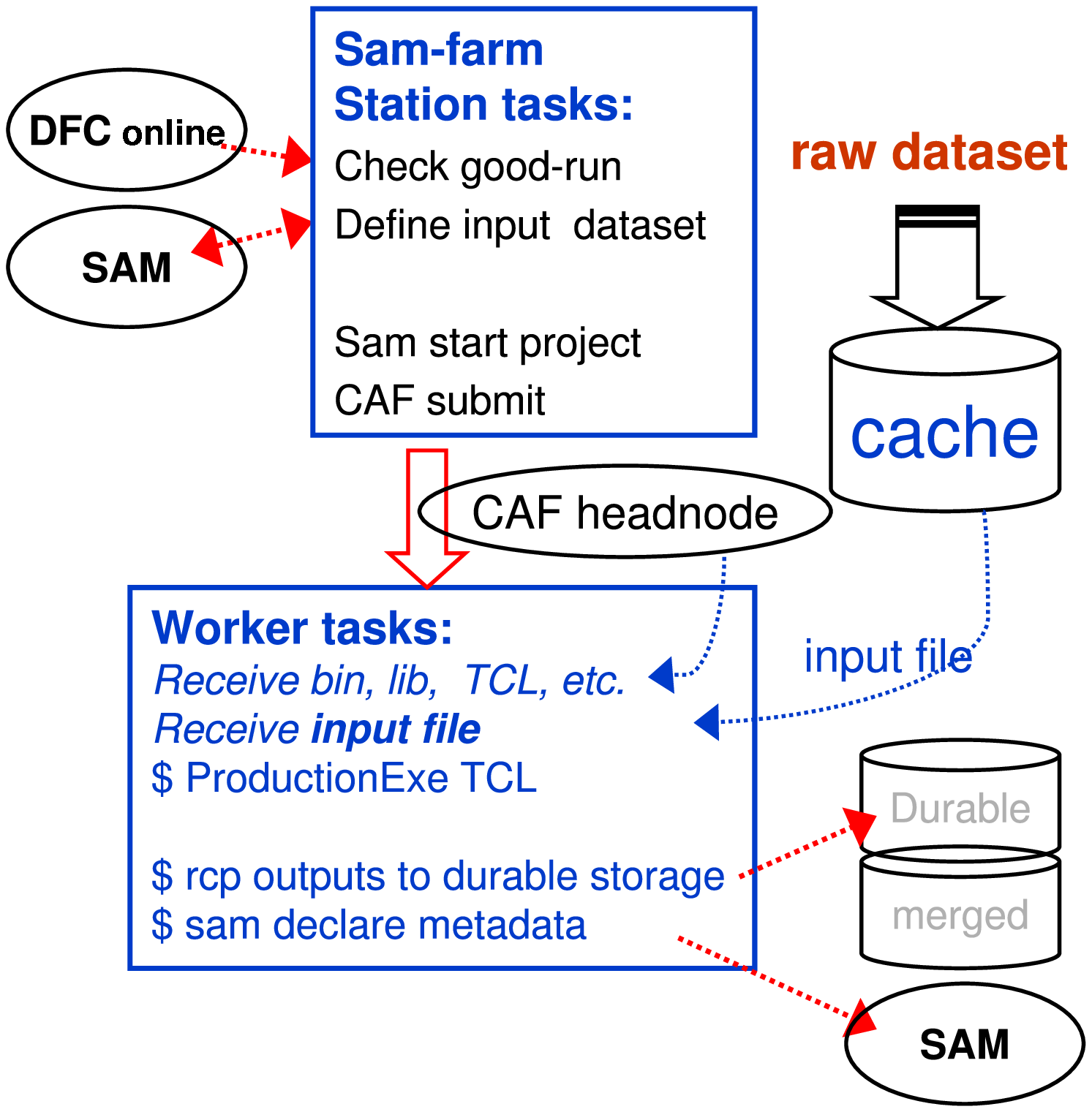,width=0.8\linewidth}
  \caption{Task flow for a SAM project submitted to CAF workers.
    A worker node receives the executable tarball and
    input data file, does the binary processing,
    then copies output to durable storage and declares SAM metadata. 
  \label{fig:cronSub} }
\end{figure}

\begin{figure}[b!]
  \centering\epsfig{file=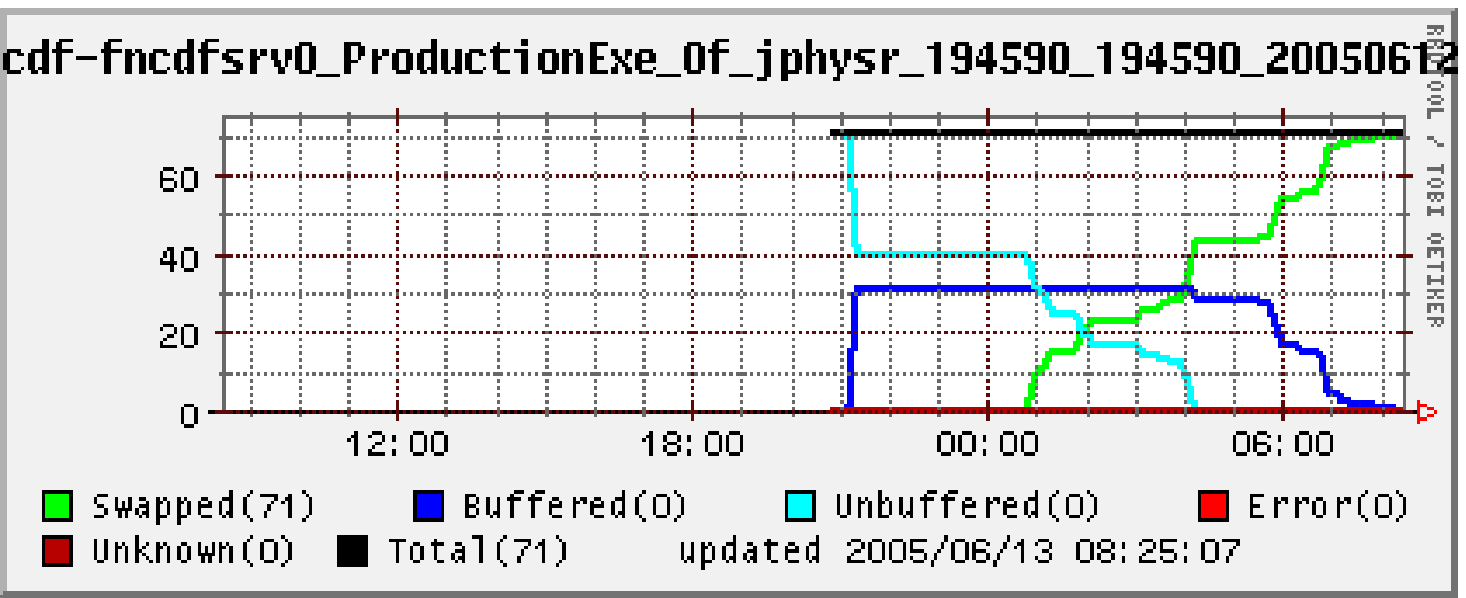,width=.90\linewidth}
  \caption{Consumption of files by a SAM project is plotted.
    Tho total of 71 files in a dataset is requested.
    Files are quickly "buffered" to CAF workers. The CAF job
    is configured to use 30 CPU segments.  After approximately
    4 hours, consumed files are being "swapped".
    The project is terminated after all files are swapped.
  \label{fig:project} }
\end{figure}

\section{SAM production farm }  

An upgrade of the production farm was required
for the increasing demand in computing capacity.
The FPS system, because of its complexity,
has become more difficult to be compatible with newly 
developed computing facilities.
Its application is converted for 
the CDF Analysis Farms (CAF) \cite{CAF}
and the SAM data handling system suitable for
distributed computing environment.

The CAF 
is a Linux PC farm with access to the CDF data management system
and databases running batch analysis jobs.
It provides software interface for job submission to batch systems 
like FBS and Condor \cite{Condor} in a uniform manner.
It is deployed in 
many CDF collaboration institutes all over the world.
The CDF data management is migrated to the SAM data handling system.
SAM is organized around a set of servers
communicating via CORBA to store and retrieve files
and associated metadata.
File information is stored in the SAM database as file metadata.
A task for processing many files is launched as a SAM project.
A project is organized for a user dataset,
with a consumer process established to receive data files.
File delivery is coordinated 
such that the events are read only once from the analyses programs
of the project.

The upgrade of production farm to a SAM based system was conducted
with minimum changes to existing hardware.
Illustrated in Fig.~\ref{fig:samfarm} is the data flow 
associated with the hardware architecture and
communication with SAM.
With the input file-caching provided by SAM, 
staging of Enstore tapes is not required.
Output files are sent to a ``durable storage'' on file servers.
These files are registered to SAM, yet short-lived
waiting for concatenation to be conducted on the file servers.

The communication with SAM database is conducted by the server nodes
configured as SAM stations.
The CAF and the durable storage nodes are entities easily specified 
in the job submission,
therefore it is flexible enough to use any facility accessible.
To improve bandwidth and file usage,
the SAM production farm is configured for direct access to the 
dCache \cite{dcache} file system where input files are 
downloaded from Enstore.
Concatenated output files are transferred directly to Enstore.

Job submission is controlled by applications scheduled on a SAM 
station.
The file metadata is also used for bookkeeping purpose.
The tasks preparing input datasets and data processing 
on a CAF worker node are illustrated in Fig.~\ref{fig:cronSub}.
The tasks are:
\begin{itemize}
\item  {\bf Prepare input datasets : } \\
       Input data to be processed are selected by
       queries to online DFC records for data of good quality (good-run)
       and detector calibration.
       The input datasets are organized in run sequence 
       of one or multiple runs for a raw data stream.

\item  {\bf Start SAM project, and CAF submission : } \\
       A SAM project is started for a dataset defined and not yet
       fully consumed.
       It is submitted to a CAF.
       SAM establishes a consumer process to deliver files 
       to CAF workers.
       From the CAF headnode workers receive an archived file (tarball)
       containing program binary, library and control parameter files.
       Input files are copied to the local scratch area.
       Files are delivered according to the file consumption status,
       until all files are delivered.
       Output files of the program are then copied to dedicated
       durable storage nodes, and the associated metadata are declared
       to SAM.
\end{itemize}

The dataset preparation and job submission are all issued
periodically by cron jobs.  
A project monitoring graph on the consumption 
of data files are plotted in Fig.~\ref{fig:project}

In comparison with the FPS system, the SAM farm management
deals with datasets.
Tracking of individual files is taken care by the SAM consumer process.
The operation is therefore reduced to detecting incomplete projects
and debugging.  The bookkeeping task is reduced from tracking thousands
of files to tracking a few dozens of projects.
The monitoring is concentrated
on the usage of durable storage, where output from CAF are checked
and merged in the concatenation process.

\begin{figure}[b!]
  \centering\epsfig{file=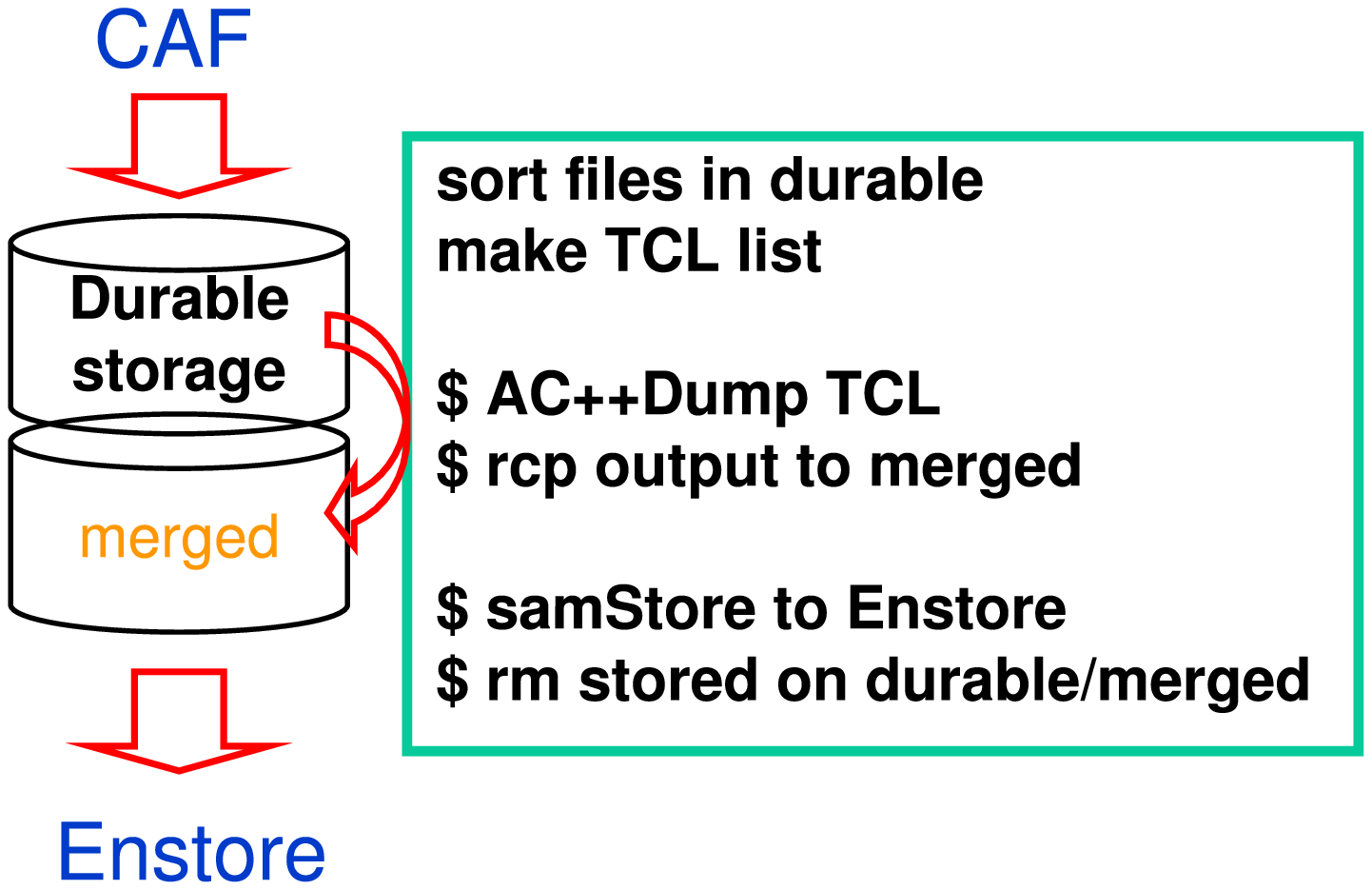,width=.80\linewidth}
  \caption{
    Files in a durable cache are sorted into lists in
    input parameter file (in TCL language) 
    read by the concatenation binary (AC++Dump).
    The merged files (of size close to 1 GB) are stored to SAM.
  \label{fig:durable} }
\end{figure}

\section {Durable storage } 

Output of CAF jobs are buffered in durable storage
on 2 TByte file servers.
When the total size of files exceeds a threshold 
(for example, 20 GByte),
a concatenation job is launched merging small files into output
of size close to 1 GByte.
Previously in the FPS system, the output of concatenation 
is truncated into 1 GByte.
Therefore an input file can be written into two concatenated files.  
This algorithm was changed to be more flexible
for merging a complete set of files.
This has simplified bookkeeping on parentage records in metadata
for unique correspondence of input and output.
The details of concatenation on the durable storage node
are illustrated in Fig.~\ref{fig:durable},
and are described in the following:
\begin{itemize}
\item  {\bf Durable cache :} \\
       The durable cache is a directory on a large file server
       where CAF output of the same dataset are stored.
       In total 43 directories are used for all reconstructed datasets.
       The files are buffered up  to a threshold (for example 100 files).
       A cron job sorts them into lists of files ordered by run.
       The number of files in a list is collected to the desired 
       concatenation file size.  And the control parameters are prepared 
       to include these files to the concatenation binary (AC++Dump).

\item  {\bf Concatenation :} \\
       Concatenation is conducted on the file server 
       and the output is stored in the "merged" directory ready
       to be stored to SAM.

\item  {\bf SAM store :} \\
       Merged files are scanned periodically to check if they exceed a
       a threshold (for example 10 GByte)
       and SAM store is conducted to copy files to Enstore
       and declare metadata.
       The threshold size is tuned to reduce Enstore operation cycles.

\end{itemize}

The concatenation job is mostly moving blocks on disks,
therefore we chose to have it processed locally on the file server
to avoid moving data on the network.
The CPU time is roughly 3 minutes per GByte on a Pentium III 2.6 GHz
file server using 7200 rpm IDE hard drives.  
While copying files to Enstore,
the network giga-link speed is commonly running at 
20 MByte/sec and the Enstore logging rate by a single mover
can accomplish over 1 TByte a day.

The new system is designed more tolerant of errors
due to hardware failure or program crashes.
The file metadata is tailored for bookkeeping 
purpose with one-to-one parentage records
and the status in process.
If a merged output file should be reprocessed,
we check out its parents for recovery.

\section{Scalability } 

The FPS system uses dfarm file system which is the collection
of IDE hard disks on workers.
With a total 200 workers, the chance of losing a file increased
whenever a worker is not accessible.
The load on MySQL database also required faster CPU for 
processing thousands of queries in an instance.
The architecture of the FPS system is restricted 
to direct data access to the Enstore.
This feature has prohibited usage other than the dedicated 
production operation.

The SAM production farm exploits the advantages of the
data handling system provided.
The usage of file metadata is convenient for bookkeeping.
Its configuration can be easily modified.
Jobs can be dispatched to any CAF facility.
And the concatenation nodes can also be 
located anywhere accessed by the CDF data handling system.
The prototype SAM production farm was tested with a SAM station
at Fermilab and jobs submitted to 
CAF facilities in Japan and Taiwan. We were able to accomplish
a few MByte/sec bandwidth.

The dedicated SAM production farm was constructed in the spring 2005
at Fermilab.  It has gigabit network links with a CAF 
of 70 workers and four file servers.
The data input is configured for direct copy from
a dCache read pool.
Each file server running two concatenation jobs
can provide a 0.5 TByte throughput rate per day.
This system has accomplished a stable operation for 
CDF data collected in 2005.
By increasing worker nodes and file servers, we expect to 
accommodate and scale beyond the 2 TByte daily processing rate 
to the maximum bandwidth and Enstore tape capacity.

\section{Conclusion}

The CDF production farms have been successfully prototyped
and commissioned.
They have provided the computing capacity
required for the CDF experiment in Run II. 
The system has been modified and enhanced during the years of 
its operation to adjust to new requirements and to enable 
new capabilities.  
The production facility is recently upgraded to adapt to the SAM data 
handling system.
It was migrated from a customized central computing model to 
a portable system for operation on distributed computing facilities.
The system will continue to be modified for higher data throughput capacity.
These developments will allow CDF to continue to process and analyze data 
through the end of the 
life of the experiment.

{}

\end{document}